\documentclass[12pt]{article}
 \usepackage{epsfig}
 \def\be{\begin{equation}}
 \def\ee{\end{equation}}
 \def\bea{\begin{eqnarray}}
 \def\eea{\end{eqnarray}}
 \usepackage{graphicx}% Include figure files

 \catcode`\@=11
 \def\lsim{\mathrel{\mathpalette\@versim<}}
 \def\gsim{\mathrel{\mathpalette\@versim>}}
 \def\@versim#1#2{\vcenter{\offinterlineskip
 \ialign{$\m@th#1\hfil##\hfil$\crcr#2\crcr\sim\crcr } }}
 \catcode`\@=12

 \catcode`@=12
\begin{document}
 \thispagestyle{empty}
 \begin{flushright}
 UCRHEP-T631\\
 July 2025\
 \end{flushright}
 \vspace{0.6in}
 \begin{center}
 {\LARGE \bf MonoHiggsology \\}
 \vspace{1.5in}
 {\bf Ernest Ma\\}
 \vspace{0.1in}
{\sl Department of Physics and Astronomy,\\ 
University of California, Riverside, California 92521, USA\\}
 \vspace{1.2in}

%{\it (in memory of Eileen)\\}
\end{center}

\begin{abstract}\
Flavor symmetry and other ideas beyond the standard model (SM) may be achieved 
in a renormalizable theory using the dark sector, while keeping only the one 
SM Higgs doublet.
\end{abstract}

\vspace{1.5in}
\noindent
Invited Contribution to 40th Anniversary of IJMPA and MPLA  
({\it in memory of Eileen}).
\newpage

\baselineskip 24pt
\noindent \underline{\it Introduction}~:~ 
On Jule 4, 2012, Peter Higgs came to CERN.  There he met Francois Englert for 
the first time.  Although they were among the recipients of the Wolf Prize 
in 2004 and the Sakurai Prize in 2010, Higgs was absent at both occasions. 
The CERN event was of course the official announcement of the discovery of 
the Higgs boson at a mass of 125 GeV, predicted by Higgs as the physical 
manifestation of the Brout-Englert-Higgs mechanism for spontaneous breaking 
of the Salam-Weinberg $SU(2) \times U(1)$ gauge symmetry of the standard 
model (SM). All its other particles, i.e. the photon, $W^\pm$, $Z$, gluons, 
as well as three families of quarks and leptons are already known. The SM 
appears to be complete.  Is there anything beyond?
\begin{table}[tbh]
\centering
\begin{tabular}{|c|c|c|c|}
\hline
particle & $SU(3)$ & $SU(2)$ & $U(1)$ \\
\hline
$(u,d)_L$ & 3 & 2 & 1/6 \\ 
$u_R,d_R$ & 3 & 1 & $2/3,-1/3$ \\ 
$(\nu,e)_L$ & 1 & 2 & $-1/2$ \\ 
$e_R$ & 1 & 1 & $-1$ \\ 
\hline
$(\phi^+,\phi^0)$ & 1 & 2 & 1/2 \\ 
\hline
\end{tabular}
\caption{Standard Model particles.}
\end{table}

There are of course two obvious missing ingredients, i.e. neutrino mass 
and dark matter.  Each offers a bewildering myriad of theoretical 
possibilities.  In this paper, a guiding principle is proposed, namely 
that there is only the one SM Higgs doublet, and all ideas beyond the SM 
are implemented through the dark sector in a renormalizable theory. This 
is thus a generalization of the scotogenic mechanism~\cite{m06,t96} 
first applied to radiative neutrino masses from dark matter.  Instead of 
using dark singlet fermions~\cite{m14}, dark singlet scalars are proposed. 
Explicit soft symmetry breaking will play a crucial role.

\noindent \underline{\it Dark Scalar Flavons}~:~ 
In the SM, all fermion masses (except possibly those of the neutrinos) 
come from Yukawa couplings of the one Higgs doublet, i.e. 
\begin{equation}
y_f \bar{f}_L f_R \left(v + {h \over \sqrt{2}}\right) = m_f \bar{f}_L f_R 
\left(1 + {h \over v\sqrt{2}}\right).
\end{equation}
The wide range of observed quark and lepton masses is a puzzle which 
prompted the idea of flavons $\eta$ which are scalar singlets carrying 
global $U(1)$ charges under which the quarks and leptons also transform, 
known widely as the Froggatt-Nielsen (FN) mechanism~\cite{fn79}.
Yukawa couplings are now of the form
\begin{equation}
y_f \bar{f}_L f_R \left(v + {h \over \sqrt{2}}\right) \left[ {\langle \eta 
\rangle \over \Lambda}\right]^k,
\end{equation}
which is nonrenormalizable (unless $k=0$).  This idea assumes some large 
cutoff scale $\Lambda$ which is a common practice of effective field 
theory, and that $\eta$ has a nonzero vacuum expectation value (VEV) 
which breaks the assumed global $U(1)$ symmetry.  The hierarchy of fermion 
masses is then correlated with the power index $k$.  Numerous implementations 
and extensions of this mechanism have appeared over the years. 

In this paper, instead of blaming our ignorance on unknown physics at higher 
energy scales, it is proposed that only renormalizable interactions are 
considered.  The key assumption is the existence of a dark sector where the 
flavons now live, and the FN symmetry is broken by explicit soft terms 
instead of spontaneously.  The effective Yukawa couplings are generated 
in one loop through the dark sector, using only the one SM Higgs doublet. 

\noindent \underline{\it Scotogenic Fermion Masses}~:~ 
\begin{figure}[htb]
\vspace* {-3.5cm}
\hspace*{-3cm}
\includegraphics[scale=1.0]{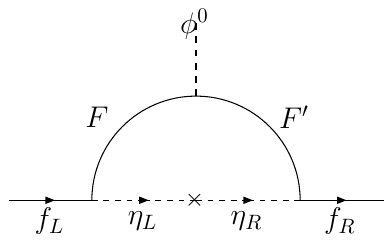}
\vspace{-22.5cm}
\caption{Generic scotogenic fermion mass.}
\end{figure}
The generic one-loop diagram for scotogenic fermion masses is shown in Fig.~1.  
The particles in the loop belong to the dark sector.  The SM fermions $f$ are 
assumed to transform under a symmetry ${\cal S}$ which may or may not allow 
them to acquire mass through the one SM Higgs doublet $\Phi$ which is a 
singlet under ${\cal S}$.  For those SM fermions which do not couple 
directly to $\Phi$, they do so in one loop through the corresponding 
heavy dark fermions, $F$ which is part of a vectorlike doublet under the SM 
$SU(2) \times U(1)$ and $F'$ which is a singlet.  They are singlets under 
${\cal S}$.  Hence  $\eta_L$ transforms in the same way as $f_L$ under 
${\cal S}$ and $\eta_R$ as $f_R$.  The soft breaking of ${\cal S}$ occurs 
through $\eta_L-\eta_R$ mixing and guarantees the loop diagram to be finite.
For an explicit example of this scotogenic realization of the Froggatt-Nielsen 
mechanism and a revised version of the original $A_4$ model~\cite{mr01} of 
leptons, see Ref.~\cite{m25}.

\noindent \underline{\it Anomalous Higgs Couplings}~:~ 
In the SM, the one Higgs boson $h$ couples to a given fermion $f$ with the 
strength $m_f/v\sqrt{2}$, where $v = \langle \phi^0 \rangle = 174$ GeV.  If 
$m_f$ is radiative in origin, then this is no longer the case, as first 
pointed out in Ref.~\cite{fm14}.  In the present framework, consider the 
$2 \times 2$ mass matrix linking $(\bar{F}_L,\bar{F}'_L)$ to $(F_F,F'_R)$, 
and choose the simplified form
\begin{equation}
{\cal M}_{FF'} = \pmatrix{M & m \cr m & M},
\end{equation}
where $M$ is an invariant mass and $m$ comes from $v$. The Yukawa coupling 
matrix of $h$ is
\begin{equation}
{\cal Y}_{FF'} = {1 \over v\sqrt{2}} \pmatrix{0 & m \cr m & 0}.
\end{equation}
The fermion mass eigenstates are now $F_{1,2} = (F \pm F')/\sqrt{2}$ with mass 
eigenvalues $M_{1,2} = M \pm m$, and the Yukawa coupling matrix in this 
basis becomes 
\begin{equation}
{\cal Y}_{FF'} = {1 \over v\sqrt{2}} \pmatrix{m & 0 \cr 0 & -m}.
\end{equation}

To complete the radiative mass calculation of Fig.~1, let $\eta_R=\eta_L^*$ and 
assume its real and imaginary parts to have the masses $M_{1,2}$ also as an 
example. Then
\begin{equation}
m_f = {y_1 y_2 (M_1+M_2) \over 64 \pi^2} \left( 1 - {M_1M_2 \ln(M_1^2/M_2^2) 
\over (M_1^2-M_2^2)} \right),
\end{equation}
whereas the $h$ coupling to $\bar{f}f$ through Fig.~1 is
\begin{equation}
y_f = {y_1 y_2 m \over 64 \pi^2 v\sqrt{2}} \left( \ln(M_1^2/M_2^2) - 
{2(M_1^2+M_2^2) \over (M_1^2-M_2^2)} + {4M_1^2 M_2^2 \ln(M_1^2/M_2^2) \over 
(M_1^2-M_2^2)^2} \right).
\end{equation}
Hence $m_f/y_f$ is not exactly equal to $v\sqrt{2}$ although it reaches 
this limit for $m << M_{1,2}$.  Even so, there are additional couplings of 
$h$ to $\eta$ which will change this result.

\noindent \underline{\it Peccei-Quinn Origin of the Dark Symmetry}~:~ 
So far, the dark fermions and scalars are simply assumed to be odd under a 
discrete $Z_2$ symmetry.  What is its origin?  Supersymmetry with $R$ parity 
is one possible answer, but it is too restrictive as to its particle 
content. A much better choice is the anomalous Peccei-Quinn global $U(1)$ 
symmetry~\cite{pq77}, the spontaneous breaking of which generates a 
dynamical axion~\cite{kc10} and relaxes the $CP$ nonconserving phase of 
quantum chromodynamics (QCD) to zero. A residual $Z_2$ discrete symmetry 
always appears in this case, as pointed out in Ref.~\cite{dmt14}.  Hence 
$F,F',\eta_L,\eta_R$ may be assumed charged under $U(1)_{PQ}$.  As long as 
there are heavy dark quarks, a QCD axion appears and may be a component 
of dark matter, in addition to the lightest particle odd under $Z_2$. For 
a recent application using the flavor symmetry $Z_3 \times Z_3$ for quarks 
and $A_4 \times A_4$ for leptons, see Ref.~\cite{m25-pq}.

\noindent \underline{\it Universal $Z_4$ Symmetry}~:~ 
In any renormalizable quantum field theory, the polynomial Lagrangian 
has terms up to dimension four only.  The quartic terms obey an universal 
$Z_4$ symmetry~\cite{m25-1,m25-2} under which any vector gauge boson 
$A_\mu \sim 1$, any scalar $\phi \sim -1$, any left-handed fermion 
$\psi_L \sim i$, and any right-handed fermion $\psi_R \sim -i$. The only 
quadratic terms are those of $\phi^2$ which also obey $Z_4$.  On the other 
hand, all cubic terms, i.e. 
$\phi^3, \bar{\psi}_L \psi_R, \psi_L \psi_L, \psi_R \psi_R$, break $Z_4$ to 
$Z_2$ explicitly but softly.  Hence their corresponding mass parameters 
may be assumed naturally small.  This new insight provides a possible 
clue to extensions of the SM, which has no cubic term itself before 
spontaneous symmetry breaking with $Z_4 = (i)^{(B-L+4Y)}$.

\noindent \underline{\it Dark $U(1)_D$ Gauge Symmetry}~:~ 
The SM has only one mass scale, i.e. the Higgs mass $m_h$ or VEV, 
$\langle \phi^0 \rangle = 174$ GeV.  In the proposed scotogenic extensions, 
possible new mass scales do appear.  The neutral scalar flavons belong in 
the dark sector and have no VEV.  Their masses should also not be much greater 
than $m_h$ to avoid large radiative corrections to the latter.  The required 
dark quarks and leptons should have large enough masses to avoid production 
at the Large Hadron Collider (LHC). Hence gauge-invariant masses of order 
TeV are called for in the form of $\bar{\psi}_L \psi_R$.  Normally, this is 
accepted without a second thought, but in view of the universal $Z_4$ 
symmetry, such mass terms should be small instead of large.

To escape this conundrum, consider a dark $U(1)_D$ gauge symmetry with 
particles shown in Table~2.
\begin{table}[tbh]
\centering
\begin{tabular}{|c|c|c|c|c|}
\hline
particle & $SU(3)_C$ & $SU(2)$ & $U(1)_Y$ & $U(1)_D$ \\
\hline
$(a,v)_L$ & 3 & 2 & 1/6 & $-1/3$ \\ 
$(a,v)_R$ & 3 & 2 & 1/6 & $1/3$ \\ 
$a'_R,v'_R$ & 3 & 1 & $2/3,-1/3$ & $-1/3$ \\ 
$a'_L,v'_L$ & 3 & 1 & $2/3,-1/3$ & $1/3$ \\ 
\hline
$(N,E)_L$ & 1 & 2 & $-1/2$ & 1 \\ 
$(N,E)_R$ & 1 & 2 & $-1/2$ & $-1$ \\ 
$N'_R,E'_R$ & 1 & 1 & $0,-1$ & 1 \\ 
$N'_L,E'_L$ & 1 & 1 & $0,-1$ & $-1$ \\ 
\hline
$\sigma$ & 1 & 1 & 0 & 2 \\ 
$\sigma'$ & 1 & 1 & 0 & 2/3 \\ 
$\eta_L,\eta_R$ & 1 & 1 & 0 & 1 \\ 
$\eta'_L,\eta'_R$ & 1 & 1 & 0 & $-1/3$ \\ 
\hline
\end{tabular}
\caption{Dark $U(1)_D$ particles.}
\end{table}
\begin{figure}[htb]
\vspace* {-5.5cm}
\hspace*{-3cm}
\includegraphics[scale=1.0]{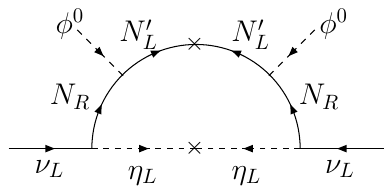}
\vspace{-21.5cm}
\caption{Generic scotogenic neutrino mass.}
\end{figure}
All these dark fermions obtain masses through $\langle \sigma \rangle$ or 
$\langle \sigma' \rangle$ and 
there are no cubic terms. It is now reasonable to have all dark masses in 
Fig.~1 of order TeV.  The scalar flavons $\eta_L \sim 1$ under $U(1)_D$, 
hence there are cubic terms $\sigma^* \eta_L \eta_L$, etc. which may be assumed 
small, resulting in the generic scotogenic diagram for Majorana neutrino 
mass shown in Fig.~2.

\noindent \underline{\it Concluding Remarks}~:~ 
The idea that family structure of quarks and leptons may be due to the dark 
sector and supported by just the one Higgs doublet of the standard model 
was originally proposed in Ref.~\cite{m14}.  The difference in its 
implementation here is the use of dark scalar flavons with the recognition 
of the universal $Z_4$ symmetry~\cite{m25-1,m25-2}.  It implies the 
possible existence of a dark $U(1)_D$ gauge symmetry, and predicts 
anomalous Higgs couplings~\cite{fm14} which are experimentally 
verifiable. The SM may be complete as it is with its one and only one 
Higgs boson.  The rest is in the dark sector.

%\newpage
\baselineskip 18pt
\bibliographystyle{unsrt}

\end{document}